\documentclass{article}

\usepackage{arxiv}

\usepackage[utf8]{inputenc} 
\usepackage[T1]{fontenc}    
\usepackage{hyperref}       
\usepackage{url}            
\usepackage{booktabs}       
\usepackage{amsfonts}       
\usepackage{nicefrac}       
\usepackage{microtype}      
\usepackage{lipsum}		
\usepackage{graphicx}
\usepackage[numbers]{natbib} 
\usepackage{doi}

\usepackage{todonotes}

\usepackage{multicol}

\AtBeginDocument{%
  \providecommand\BibTeX{{%
    \normalfont B\kern-0.5em{\scshape i\kern-0.25em b}\kern-0.8em\TeX}}}

\usepackage{listings}
\usepackage{xcolor}

\definecolor{codegreen}{rgb}{0,0.6,0}
\definecolor{codegray}{rgb}{0.5,0.5,0.5}
\definecolor{codepurple}{rgb}{0.58,0,0.82}
\definecolor{backcolour}{rgb}{0.95,0.95,0.92}

\lstdefinestyle{mystyle}{
    backgroundcolor=\color{backcolour},   
    commentstyle=\color{codegreen},
    keywordstyle=\color{magenta},
    numberstyle=\tiny\color{codegray},
    stringstyle=\color{codepurple},
    basicstyle=\ttfamily\footnotesize,
    breakatwhitespace=false,         
    breaklines=true,                 
    captionpos=b,                    
    keepspaces=true,                 
    numbers=left,                    
    numbersep=5pt,                  
    showspaces=false,                
    showstringspaces=false,
    showtabs=false,                  
    tabsize=2
}

\title{CircleChain: Tokenizing Products with a Role-based Scheme for a Circular Economy}

\author{ Mojtaba Eshghie\\
	KTH Royal Institute of Technology\\
    Stockholm, Sweden  \\
	\texttt{eshghie@kth.se} \\
	\And
	Li Quan\\
	University of Copenhagen\\
    Copenhagen, Denmark  \\
	\texttt{lq@di.ku.dk} \\
	\And
	 Gustav Andersson Kasche\\
	KTH Royal Institute of Technology\\
    Stockholm, Sweden  \\
	\texttt{gustavak@kth.se} \\
	\And
	 Filip Jacobson\\
	KTH Royal Institute of Technology\\
    Stockholm, Sweden  \\
	\texttt{filipjac@kth.se} \\
	\And
	Cosimo Bassi\\
	Algorand Inc.\\
     Trani, Italy  \\
	\texttt{cosimo.bassi@algorand.com} \\
	\And
	Cyrille Artho\\
	KTH Royal Institute of Technology\\
    Stockholm, Sweden  \\
	\texttt{artho@kth.se} \\
}

\renewcommand{\shorttitle}{}

\hypersetup{
pdftitle={A template for the arxiv style},
pdfsubject={q-bio.NC, q-bio.QM},
pdfauthor={David S.~Hippocampus, Elias D.~Striatum},
pdfkeywords={First keyword, Second keyword, More},
}
\begin{document}
\maketitle
\begin{abstract}
In a circular economy, tracking the flow of second-life components for quality control is critical. Tokenization can enhance the transparency of the flow of second-life components. However, simple tokenization does not correspond to real economic models and lacks the ability to finely manage complex business processes.
In particular, existing systems have to take into account the different roles of the parties in the supply chain. Based on the Algorand blockchain, we propose a role-based token management scheme, which can achieve authentication, synthesis, circulation, and reuse of these second-life components in a trustless environment. The proposed scheme not only achieves fine-grained and scalable second-life component management, but also enables on-chain trading, subsidies, and green-bond issuance. Furthermore, we implemented and performed scalability tests for the proposed architecture on Algorand blockchain using its smart contracts and Algorand Standard Assets (ASA). The open-source implementation, tests, along with results are available on our Github\footnote{https://github.com/Kasche153/CircleChain} page.

\end{abstract}


\section{Introduction}

\subsection{A Case for E-Waste}\label{subsection:ewaste}
The total number of electronic devices globally, such as mobile devices, is currently more than the number of humans on earth and is increasing \cite{mnumber}. Accordingly, the waste generated by discarding these devices is also staggering. Based on Global E-waste Monitor (GEM) 2020 report, global e-waste generation is 53.6 Mt or 7.3 kg per capita \cite{gem2020} with an annual growth rate of 2 Mt. Only 17.4\% ($9.3$ Mt) is formally documented as collected and recycled. This shows recycling activities are neither currently sufficient nor keeping pace with the global e-waste generation increase. Because of the high volume in their generation and the possibility of collecting, reusing, and refurbishing them, providing a system that uses a circular economy (section \ref{subsection:circular-economy}) for e-waste would be a significant contribution to a sustainable future. We will use the supply chains of electronic chips as an example from the e-waste category to demonstrate how our system works.

\subsection{Circular Economy}\label{subsection:circular-economy}
Reusing and recycling products and product parts has environmental and socio-economic benefits. The closed-loop of reuse and recycling that aims to minimize resource consumption is called circular economy (CE) \cite{targetsforce2020}, which contrasts with the linear supply chain model of taking, making, and disposing of.  Incorporating CE benefits sustainable development while reducing undesirable consequences of overuse of resources \cite{babbittClosingLoopCircular2018, miliosPoliciesResourceEfficient2016}. This is also part of the United Nations' 2030 Agenda regarding sustainable development \cite{TransformingOurWorld}.\\
Although the linear model is deeply standardized in organizations, the need for more sustainable supply chains demands abandoning it for CE \cite{hartleyPoliciesTransitioningCircular2020}. Currently, only $8.6\%$ of the worldwide economy is circular \cite{CGR2022}. According to the circularity gap report, \cite{CircularityGapReporting} this number is steady in the 4 recent years. This shows the need for platforms and solutions that facilitate CE. Based on previous research on second-life components management (part of CE) reviewed in Section~\ref{relatedwork}, blockchain is one of these CE enablers.

The Covid-19 pandemic caused material and resource shortages  in supply chains \cite{queirozImpactsEpidemicOutbreaks2020}. Santosh et al.~\cite{nandiRedesigningSupplyChains2021} argue that using a blockchain to manage second-life components in a circular economy increases supply chain agility and localization. Localization is known as designing, producing, or delivering through in-house technology and features. Agility is an organization's ability to actively detect and respond to supply chain risks. \\
We argue that a blockchain-based system for CE needs to be backed by a corresponding blockchain architecture where the cost and complexity are proportional in relation to the products or components. In other words, any blockchain-based solution used to facilitate a circular economy has to be justifiable based on carbon footprint, cost-per-transaction, and complexity of its operation.



\subsection{Tokenization}
Tokenization is the process of digitizing any type of asset on a blockchain.
By assigning a token to a product or component, its life cycle can be traced entirely and transparently. This tracking is \emph{trustless} in that all the data storage and computation take place on a shared blockchain without requiring a centralized authority.

Tokens can be fungible (for stateless components that can be replaced with an equivalent) or non-fungible (for components or products where provenance or traceability matters, such as memory chips, artwork, or jewelry). In both cases, tokens are associated with metadata containing additional information about them (such as the chip type) and stored off-chain but authenticated by a hash value stored on the blockchain.

Tokenization increases traceability, thus, making the ownership and pollution management of the components more accessible. It also makes it possible to keep a history of the components, which helps obtain information about products' \textit{composition}, \textit{ownership}, and \textit{status}. Essentially, we can issue \textit{material passports} based on this technology. Specifically, tokenization allows to trace where they are going for second-life components and if the product that is working as part of a system is indeed produced in company X and has performed only Y months. Therefore, after scrapping the original product, its constituents will not lose their intrinsic value and might be used directly in other products \cite{TokenizationBchain}.
In addition, since the reuse of components previously used in other consumer products (second-life components) is an integral part of CE, we provide a reference architecture and its implementation on the Algorand blockchain to be used as a blueprint to implement CE.

\subsection{Related Work}
\label{relatedwork}
Research on real-world solutions for handling second-life components has remained limited to theoretical backgrounds of supply chains, and there is no functional reference architecture. Most of research is limited to realizing the potential of blockchain and smart contracts to \textit{track and trace} the assets \cite{cupiInternetValueCircular2022, heimDressCodeDigital2022, Alexandris_Katos_Alexaki_Hatzivasilis_2018}.
Bekrar et al.~\cite{bekrarDigitalizingClosingoftheLoopSupply2021} evaluate the opportunities, trends, and potential uses of the Nexus blockchain \cite{UniversalBlockchain} in the context of CE by considering the transportation aspect of CE without proposing neither a solution nor a design based on smart contracts.

Morrow and Zarrebini \cite{TokenizationIndividual2019} explore use cases of tokenization for assets based on their societal implications. This study emphasizes the pivotal role that blockchain can play in the sustainability of tires.

Research on waste management via blockchain digitization and governance gives a perspective on a blockchain-based solution for handling second-life components. França et al.~\cite{francaProposingUseBlockchain2020a} propose incorporating a blockchain-based \textit{green coins} system into an already existing mobile application used to collect and sell waste and receive points that are convertible to new products in local stores. Although these works constitute a practical approach to part of the supply chain management problem we face while studying CE, their research does not include the architecture or design of a system providing guidelines for implementation.
\\
The design proposed by Hatzivasilis et al.  \cite{greenblockchainsce} includes an architecture with three roles \textit{asset operators}, \textit{asset auditors}, and \textit{regulators}. The counterpart roles are defined in our system\ref{section:role-based}. However, the \textit{asset auditor} and \textit{regulator} are both defined as \textit{authenticators} in our system and \textit{operators} are the same as \textit{users} in our system. Furthermore, this work proposes to build the system on top of the Ethereum blockchain. But, it does not provide an implementation.


Hatzivasilis et al. \cite{ANASTASIADIS2022196} performed a holistic end-to-end study on the role of traceability systems in the supply chains of the food sector. They found that the adaptation of traceability systems has a positive effect on the transition to a more CE. 

Shojaei et al. \cite{builtsector} provide proof-of-concept of a system to enable CE in the construction sector. They suggest using a permissioned blockchain system rather than public blockchains.

The most relevant work to the current effort was the blockchain-based supply chain traceability of COVID-19 \cite{COVID-19}. They implemented a supply chain tracing system using smart contracts on the Ethereum blockchain. Although their design is product specific, it could be used to solve traceability challenges in other domains. This study concludes that using smart contracts can reduce information asymmetry and other inefficiencies in supply chain tracing. The cost and security evaluation of the system showed that it was feasible as a real-world application. This study also highlights some limitations to their system one of them being the throughput of the Ethereum network which is around tens of transactions per second and limits the scalability of the system~\cite{COVID-19}. Another limitation is the time delay of the Ethereum blockchain. The confirmation time varies from block to block because of the PoW consensus algorithm.

\subsection{CircleChain}
As suggested in the section \ref{relatedwork}, the possibility of using blockchain-based architecture for a CE is well-studied. However, there are very few practical designed systems along with a working implementation that is suitable for real-world use-cases. Furthermore, the studies that try to provide an architecture either use obsolete technology or use systems that defeat the purpose of sustainability. Therefore, first, we gathered a list of requirements to choose the underlying blockchain system that is presented in section \ref{reqs}. Next, we designed a reference role-based architecture that complies with the gathered requirements, and we implemented it on a sustainable blockchain system. Finally, we tested it in a synthetic scenario to demonstrate the scalability of our solution for a real-world large-scale circular economy. 

\section{Tokenized Circular Economy Using Blockchain}
\label{reqs}
We identify the following characteristics required for a blockchain platform in the context of tokenization of products for CE:
\begin{enumerate}
    \item \MakeUppercase{Affordability}: This characteristic is with regards to development, deployment, management, and last but not least transaction cost. 
    \item \MakeUppercase{Sustainability}: The energy consumption and accordingly carbon footprint of the platform should be justifiable with the sustainability goals in mind. 
    \item \MakeUppercase{Scalability}: The chosen platform should have the capability of handling a large number of transactions in an increasingly short amount of time. 
    \item \MakeUppercase{Security}: Since successful attacks on widely known blockchain platforms are not an unknown condition \cite{chen2020survey}, the security of a platform that handles millions of products should be assured. 
\end{enumerate}

\subsection{Algorand}
Algorand is a proof-of-stake fast-consensus blockchain platform~\cite{giladAlgorandScalingByzantine2017}. Since it does not work with the concept of \textit{miners}, the consensus requires a negligible amount of computation \cite{chenAlgorandSecureEfficient2019}. 
While there are multiple other blockchain platform choices for smart contracts, the above-mentioned requirements for a CE are particularly met with Algorand.

First, each transaction in other blockchain platforms such as Ethereum \cite{wood2014ethereum} costs at least a few dollars. This transaction cost escalates depending on the logic of the smart contract. Since our design should be useful for different products ranging from a few dollars (such as electronic chips) to millions of dollars, we should base our design on a universally feasible platform.  Furthermore, this transaction cost (Gas in Ethereum blockchain) is not deterministic and it is only possible to estimate it.
Second, the carbon footprint related to each transaction in proof-of-work blockchain platforms such as Ethereum makes such blockchains a poor platform for CE~\cite{carbonfootprintbitcoin}. Third, the delay of a transaction in Ethereum is not only considerable but also non-deterministic and can take from seconds to minutes. 
Fourth, Algorand supports ASAs (see Section \ref{asasubsection}) as a Layer-1 primitive, ensuring the same security, scalability, speed, and costs of the native token (Algo) without referring to a user-defined token standard (such as ERC-20 or ERC-721 on Ethereum \cite{ERC721NonFungibleToken, ERC20TokenStandard}) that introduce complexity or even flaws~\cite{vladimirovattack}. 


\subsection{Algorand Standard Assets (ASAs)}\label{asasubsection}
Algorand Standard Assets (ASAs) are Algorand Layer-1 solutions that represent stable coins, digital art, securities, etc. \cite{WinNT}. 
Based on the aforementioned features of ASAs, it is possible to create both fungible and non-fungible tokens in only one transaction \cite{kattenIssuingGreenBonds2021}. In fact, it is possible to create an ASA using only one \textit{goal} command as in listing \ref{asagoal}.

\begin{lstlisting}[caption= {Goal command to create an ASA with a total supply of 1000, unit name of \emph{uname} and address \emph{addr}}, label={asagoal}]
goal asset create --creator addr --total 1000 --unitname uname --asseturl "https://path/" --decimals 0   -d data
\end{lstlisting}

We used ASAs to implement Non-Fungible Tokens (NFT) for chips that are used as chip passports. Sending and receiving ASA tokens requires a specific process at which the receiver has to first opt-in to the ASA. Then, the sender will be able to transfer the asset to the receiver. An opt-in transaction is an asset transfer with an amount of 0 that can be triggered using the goal command in listing \ref{optinasa}. Notice that both sender and receiver of this transaction are the receivers of ASA. 

\begin{lstlisting}[caption= {Goal command to opt-in to an ASA}, label={optinasa}]
goal asset send -a 0 --asset <asset-name>  -f <opt-in-account> -t <opt-in-account> --creator <asset-creator>  -d data
\end{lstlisting}



\section{Role-based Token Management Scheme}\label{section:role-based}

\begin{figure}
    \centering
    \includegraphics[scale=0.9]{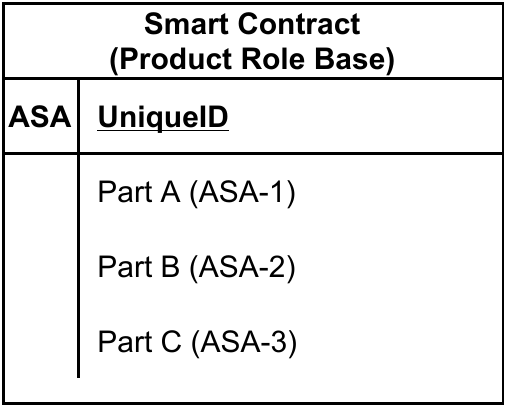}
    \caption{Two-layer architecture for role-based access to tokens}
    \label{fig:arch}
\end{figure}

Our scheme uses a two-layer architecture for role-based access management.
The tokens are managed as ASAs at the lower layer, while the role-based access management is taken care of by smart contracts that govern when state and ownership modifications in the tokens are permitted (see Figure~\ref{fig:arch}).

Our system is designed and implemented around the following roles:
\begin{enumerate}
    \item \textbf{Authenticator}
    \item \textbf{Manufacturer}
    \item \textbf{User}
    \item \textbf{Recycler}
\end{enumerate}

\begin{figure*}[h]
    \centering
    \includegraphics[width=0.9\textwidth]{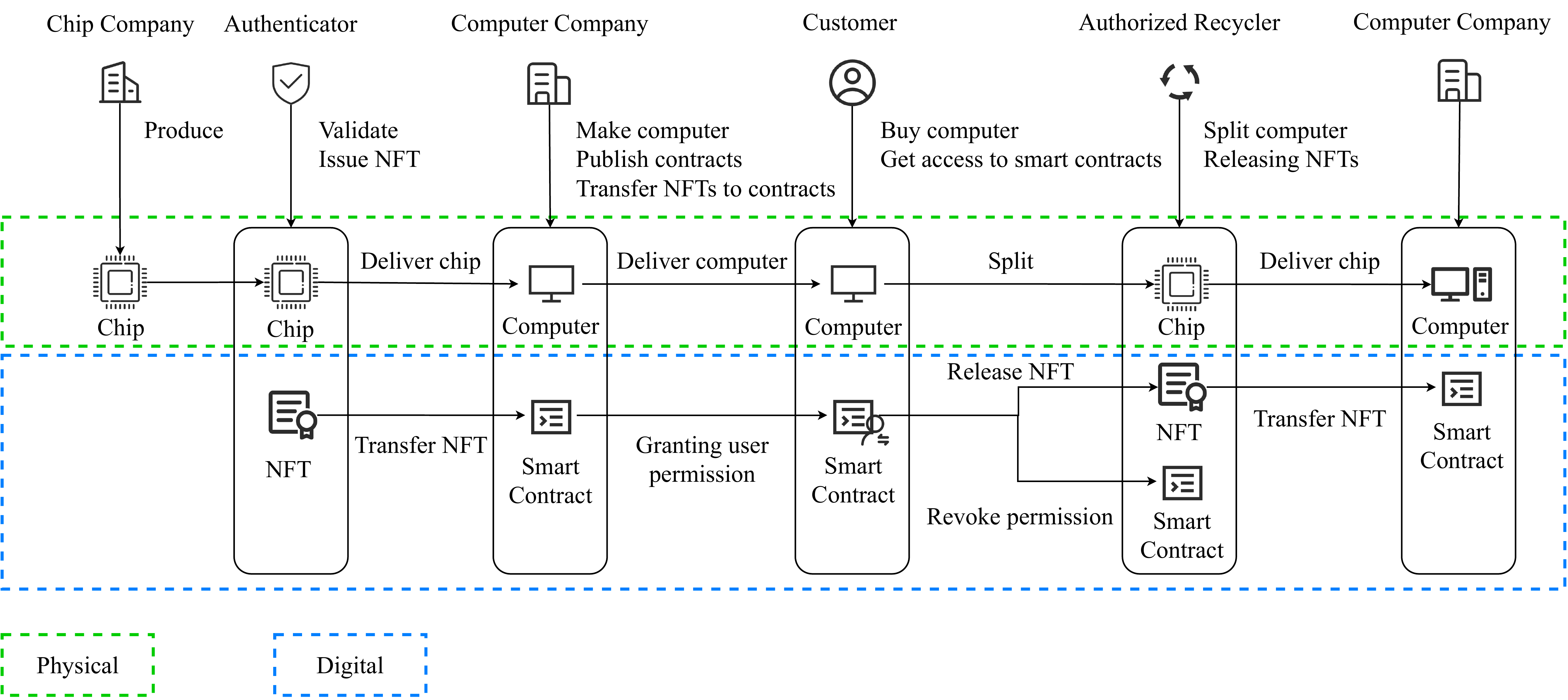}
    \vspace{-1\baselineskip}
    \caption{The life cycle of computer chips and their respective blockchain-based tokens in our system}
    \label{fig:System_Design}
\end{figure*}

\subsection{Example workflow}
The second life of chips is an important application scenario for a circular economy and sustainable development due to the worldwide shortage of chip production capacity and the opaqueness of the chip reuse market. For this reason, we demonstrate the token management scheme with the example of chips.

The second life of a chip includes repeated certification, distribution, use, and finally recycling of products using them. The participants involved include chip companies, authenticators, computer companies, users, and authorized recyclers. Different participants have different interests and authority, so they need to cooperate with the corresponding management mechanisms and process designs to ensure a fair and transparent system.

In order to accommodate the needs of the actual business process where multiple parties are involved with different authorities, we propose a role-based token management scheme. 

In Fig.~\ref{fig:System_Design}, we show the proposed role-based token management scheme by taking the second-life process of a chip as a simple example. 
The proposed process is shown in Fig.~\ref{fig:System_Design}. Specifically, the whole management mechanism consists of two parts: physical processes and on-chain operations. Transparent management of chip flow is achieved through the mapping of on-chain operations and physical processes. Compared to the primary mapping between physical components (chips) and tokens, the double-layer token management scheme adds a new token management layer (smart contract or application) to it, enabling the mapping of business processes and on-chain operations. In this mechanism, the Non-Fungible Token (NFT) represents each unique chip, and the smart contract represents the integrated product. The NFT is not delivered directly to the customer but flows between authorized organizations, such as authenticators, computer companies, or authorized recyclers. An NFT can only be issued by authenticators and released for use or destroyed by authorized recyclers, thus avoiding the current disorder and opacity of the chip's second-life market.

When a computer company produces a computer, the NFT is packaged into a smart contract that defines the roles and permissions of the chip companies, authenticators, computer companies, customers, and authorized recyclers. The authenticator is an independent third-party organization with relevant professional qualifications. This scheme provides a realistic reflection of the actual business process. For example, only the authenticator can validate the initial properties of the chip. After delivering the computer to the customer, the computer company also needs to provide after-sales service to the customer. Only authorized recyclers have the authority to handle the chips.

The management of a chip consists of six processes: production, certification, usage, distribution, and recycling. 
\begin{enumerate}
    \item Production: The chip company produces the chip. 
    \item Certification: After validating the design and manufacturing of the chip, the authenticator issues a corresponding ASA-based NFT certificate for each chip separately. These NFTs will include various information about the chip, and it is guaranteed that this information will be tamper-proof. At the same time, once these NFTs are issued, they will accompany the chip throughout its life cycle, which means that only chips bound with NFTs are considered to be authenticated chips.
    \item Usage: When the computer company purchases these chips, it is also given the management rights of the corresponding NFT. For each manufactured computer, it will come with a package that includes the NFTs of all the chips being used.
    \item Distribution: When a user acquires a computer, the user will be given the appropriate user rights in the corresponding smart contract. Note that the computer manufacturer and the chip company also have rights in this contract.
    \item When a user no longer uses her computer, she can choose to refurbish/recycle them to give chips and other components a second life by handing over their computers to authorized recyclers. 
    \item Recycling/Refurbishment: The recycler can dismantle the computer and remove the user's privileges from the smart contract, releasing the tied NFTs for future use. These chips and NFTs can be delivered to computer companies again or simply destroyed, depending on the condition of the chips.
\end{enumerate}


Our token management scheme is generic and adaptable to a variety of use cases.
Different companies and organizations can define their own token management functions based on specific business scenarios and business logic on top of the architecture we provide and still maintain the generality of the underlying tokens.


\subsection{Implementation on Algorand}


In the double-layer token management scheme, the lower layer manages the tokens, while the upper layer manages access by different entities. The lower layer is implemented using ASA-based NFT and the basic information of the chip, while the smart contract in the upper layer maintains the relationships and permissions between various entities. The Algorand blockchain provides a fine-grained token management mechanism. For instance, a computer is bonded to a Contract Account. Manager Address and Reserve Address will represent who can manage ASAs and who reserves other ASAs.

The Algorand blockchain provides two ways to implement smart contracts: (1) Smart Signature and (2) Application (smart contracts). A SmartSig is a stateless program that can control an Algorand account, based on Transaction Execution Approval Language(TEAL) logic. It can programmatically approve/reject transactions based on conditions. A SmartSig can not handle local state and dynamically change values or access an ASA's state. Contrarily, Application is a stateful program that can read and write the state of the chain, including accounts' states, ASAs' states, and global/local states. An Application is associated to an Account, which can create and submit transactions.



For example, when there are three ASAs $a, b, c$ of three chips, SmartSig achieves the recycling process with the following actions:

\begin{enumerate}
    \item SmartSig approves AssetTransfer if and only if is triggered by a known trusted entity in collaboration with the current Asset Manager.
    \item SmartSig approves if and only if $a, b, c$ are sent to Asset Manager.
    \item SmartSig approves if and only if the Product Account is destroyed in the same call.
\end{enumerate}

In the current work, we use Algorand smart contracts rather than smart signatures. 

\subsection{Components}
\begin{figure*}[h]
    \centering
    \includegraphics[width=1\textwidth]{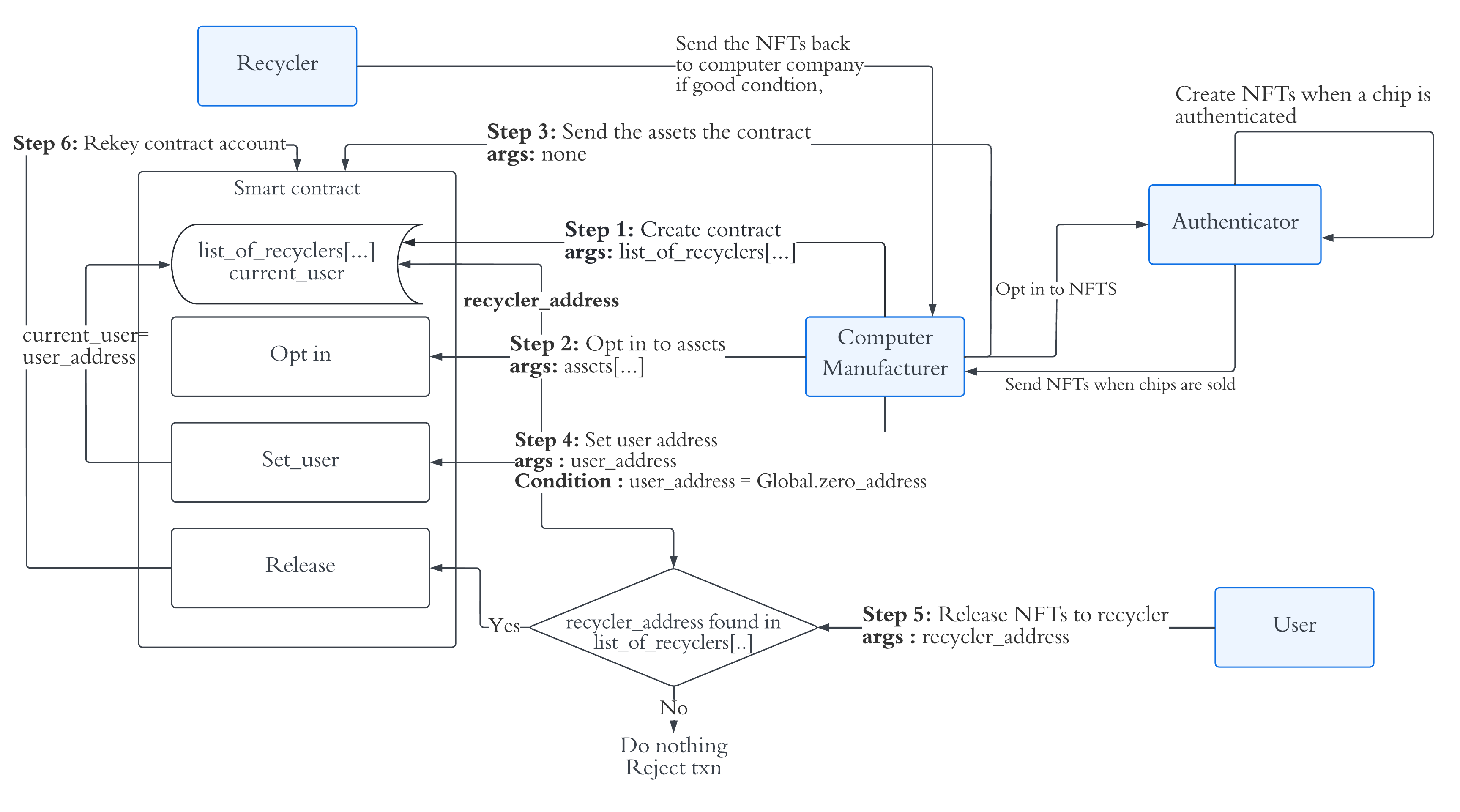}
    \vspace{-1\baselineskip}
    \caption{Workflow of system parts in CircleChain. The boxes with blue backgrounds represent roles the system is built around.}
    \label{fig:workflow}
\end{figure*}

One of the options to write the smart contracts (applications) in Algorand is PyTeal \cite{PyTealAlgorandDeveloper}. PyTeal enables writing the smart contract front-end system using py-algorand-sdk library. Without PyTeal one has to use TEAL which is Algorand bytecode language. This would make the process of writing smart contract logic harder for the developers. Besides, the notion of front-end refers to off-chain components for smart contract management.\\
Our front-end of implementation consists of \textit{creating}, \textit{compiling}, \textit{deploying}, and \textit{interacting} with a \textit{smart contract} that is the backbone of our system. Each of the operations is defined using functions in PyTeal in a \textit{contract.py} available in our repository. \\
Listing \ref{main_part_contract} shows how the smart contract is created, compiled, deployed to the blockchain, and interacted with.
\\
At line 16 (listing \ref{main_part_contract}) you can see the workflow of the system in Figure \ref{fig:workflow} is implemented to initialize, set user, or release operations using a contract call. Every call to the main application can contain a list of assets in case the logic requires them based on request parameters. \\

The complete version of the smart contract implementing the backbone logic of the architecture as well as utilities written to handle operations is available in \footnotemark[1].

\begin{lstlisting}[language=Python,label=main_part_contract,caption=Front-end usage of our Algorand smart contract system for second-life material management]
algod_client = algod.AlgodClient(algod_token, algod_address)
creator_private_key = get_private_key_from_mnemonic(user_mnemonic)
clear_state = clear_state_program()
approval = approval_program(mnemonic.to_public_key(recycler_mnemonic))

app_id = deploy_new_application(
    algod_client, creator_private_key, approval, clear_state)

app_add = logic.get_application_address(app_id)
asset_id = create_asset(creator_public_key=computer_add, creator_private_key=computer_key,
                        asset_name="CircleToken", unit_name="CT1", algod_client=algod_client, manager_public_key=computer_add,
                        total_supply=1)

algo_transaction(add=computer_add, key=computer_key,
                 reciver=app_add, amount=1000000, algod_client=algod_client)
call_contract(app_id=app_id, args="Init", assets=[asset_id],
              private_key=computer_key, public_key=computer_add)
send_asset(algod_client=algod_client, asset_id=asset_id, asset_sender=computer_add,
           asset_reciver=app_add, sender_private_key=computer_key)

call_contract(app_id=app_id, args="Set user",
              private_key=computer_key, public_key=computer_add)
call_contract(app_id=app_id, args="Release",
              public_key=user_add, private_key=user_key)
send_asset(algod_client=algod_client, asset_id=asset_id, asset_reciver=computer_add,
           asset_sender=app_add, sender_private_key=recycler_key)

\end{lstlisting}

\section{Discussion}

\subsection{Scalability of CircleChain}
One of the most important considerations in creating a blockchain-based system is its scalability. The notion of scalability depends on the context in which the system is used for. Here, it is important to have a system with high throughput for managing product tokens. \\
One potential bottleneck is where the tokens (ASAs) are sent and received. This is where opt-in operations also happen which we discussed in section \ref{asasubsection}. Here, it is possible to put as many transactions into API requests and send them. This will theoretically increase up to the block capacity. In other words, we will be able to send opt-in transactions since there is no need for grouping them for atomic execution. The practical maximum transaction per block in Aglorand is 5000 and a new block is produced every 4.5 seconds. Therefore, we can assume a throughput of 1000 transactions per second.

\subsection{Design Choices}

\subsubsection{\textbf{Algorand vs Other Blockchain Platforms}} 
The main differences between four of the most known blockchain platforms are demonstrated in table \ref{tab:blockchains}.

As also obvious from the table, most of the current blockchain systems in use do not have a deterministic transaction delay. This makes it hard to understand the behavior of the application created based on these systems at scale. For instance, our tests in section \ref{subsection:test} would not provide deterministic results if the time between two consecutive blocks in Algorand was not deterministic. \\
The other factor is the transaction cost which should be as low as possible so that building an application for supply chain management systems is justifiable. This will be of much more importance when the value of the commodity being handled is not high. In our used example, the electronic chips, this effect of transaction fee is projected because of the product value. \\
An alternative is to use a private deployment of Ethereum or any other blockchain platform which can be configured to have higher transaction throughput and redefine the transaction fee manually \cite{helliar2020permissionless}. However, such an approach would defeat the purpose of using a blockchain system in the first place since a permissioned blockchain does not have the same distributedness that a public blockchain has such as Ethereum main network. \\
Last but not the least, we need to take into account the environmental impacts our transactions have. Traditional proof-of-work (PoW) consensus systems are not environmentally friendly and should be avoided. This along with the current transaction throughput are the main reasons for not using Ethereum main network to deploy our application.

\begin{table*}[h]
  \caption{Comparison of blockchain platforms and their characteristics to be used in a large-scale circular economy. }
  \label{tab:blockchains}
  \centering
  \begin{tabular}{p{1.6cm}|p{1.4cm}|p{4.4cm}|p{3cm}|p{1.8cm}|p{1.6cm}}
     
    \toprule
    \textbf{Platform} & \textbf{Consensus} & \textbf{Considerations} & \textbf{Transaction cost\footnotemark[2]} & \textbf{Throughput\footnotemark[3]} & \textbf{TX Delay}\\
    
    \midrule
    \textbf{Algorand} \cite{chen2019algorand} & PoS & ASAs, PyTeal: Python language binding & constant  $\approx  \$0.0005$ & $\approx 1000~tps$ & 4.5 secs\\
    \midrule

    \textbf{Ethereum} \cite{wood2014ethereum} & PoW & Turing-complete and easy to use language, Vulnerabilities:reentrancy, etc., huge community of developers and experts  & variable, unbounded transaction time.  $\approx \$1.78 $  &
    $\approx 30~tps $ & 15 sec - 5 mins \\
    
    \midrule

    \textbf{Solana} \cite{yakovenko2018solana} & PoH & - & $\approx \$0.0002 $ & 65000 tps & secs - mins \\
    \midrule

    \textbf{Avalanche} \cite{AvalancheWhitepaperWhitepaper} & Avalanche & Uses EVM &  \$0.00025 & 5000~tps  &2-10 secs \\
    \bottomrule
    
  \end{tabular}
\end{table*}
    \footnotetext[2]{ The cost of the transaction is reported at the time of writing this paper.}
    \footnotetext[3]{ the throughput is measured in transactions per second (tps), and is defined as the total number of transactions issued by network participants that  the the ledger can confirm in a second.}


\subsubsection{\textbf{NFT vs FT}}
The proposed architecture can support Non-Fungible Token (NFT) and Fungible Token (FT) as essential asset representatives, depending on the type of asset. For example, for chips, each chip is unique and has a unique identification number. This makes it difficult to tamper with and fits the characteristics of NFT. In the case of products that each unit is not or does not need to be represented as unique, FT is suitable. For instance, glass as a building block of other products might fit with FT tokenization.
\\
Since ASAs act as a template for creating tokens in Algorand, changing the type of the token is as easy as changing the \textit{total} parameter in an asset configuration transaction. It is also possible to add more information to the token to \textit{unit name} and \textit{asset name} parameters based on the token type.

\subsubsection{\textbf{Authenticators}}
A big issue in reusing and recycling products is the quality of products built with second-hand materials and premature recycling. Our current system and implementation do not include a procedure to prevent these issues. However, we included an \textit{authenticator} role that keeps account of recyclers and can be effective to tackle mentioned problems. This role can be extended to further include other features as well. 

\subsubsection{\textbf{Recyclers}}
As mentioned earlier, a circular economy is defined by not just recycling, but also by refurbishing, reconditioning, repairing, and upgrading products or product parts. This means that the current role with the name \textit{recyclers} in our system can in fact have different roles in the real world other than merely scraping and recycling an otherwise reusable product. 

\subsection{Testing at Scale}\label{subsection:test}
\begin{figure}
    \centering
    \includegraphics[width=0.7\columnwidth]{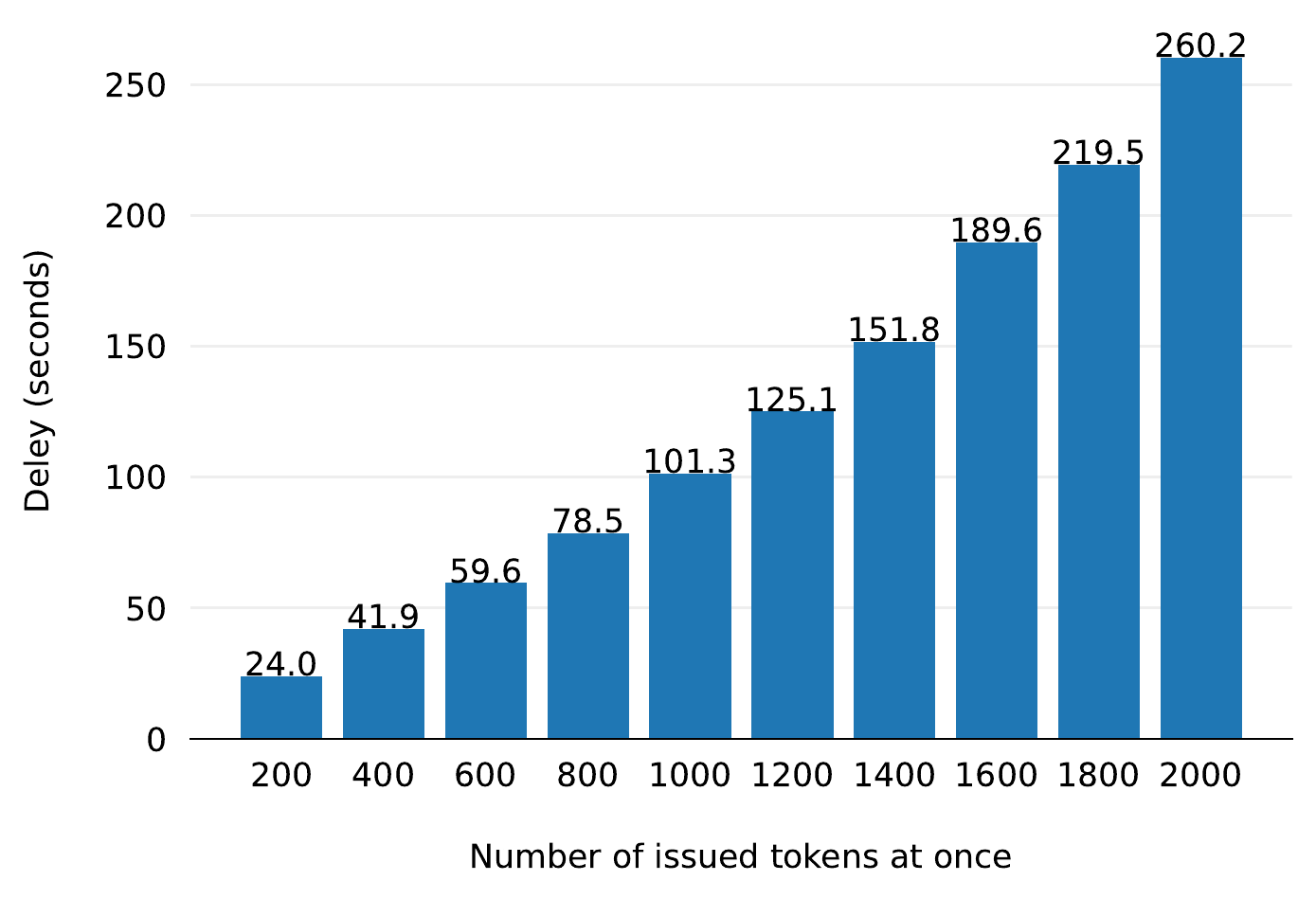}
    \caption{The time it takes to create a batch of tokens with consecutive API calls to Algorand Testnet. }
    \label{fig:timetoconfirm}
\end{figure}

As pointed out earlier our current design choice for the type of tokens is NFT which means for each product (here: chips) a new token will be issued in the form of Algorand ASA. \\
To show how this scheme will scale when large batches of tokens are issued for physical products, we issued batches of token creations and measured the time it takes for the transactions to be confirmed for all of the tokens in the batch. In the tests, we ran all token creation transactions from the same machine, and the creators of all tokens in the batch are the \textit{authenticator} role, as it is shown in figure \ref{fig:workflow}. 
\\
As the results of this experiment in figure \ref{fig:timetoconfirm} shows, the total time it takes to create a batch of tokens increases linearly with the size of the batch. This shows that increasing the scalability of the system by parallelizing the API calls to the Algorand client for token creation is very effective in keeping the asset creation time within reasonable bounds.

The environment setup where the scalability test was performed is shown in table \ref{tab:test}. The performed tests are available in our CircleChain Github repository\footnote[4]{https://github.com/Kasche153/CircleChain/tree/main/sandbox/tests}. 

\subsection{Basic Functionality}
In our present model as well as implementation the access control mechanisms that restrict access to the contract in a dynamic way are missing. In other words, the access control mechanisms currently in place only allow a defined set of accounts to be used with permissions to create the tokens. This provides the basic functionality required to operate a circular supply chain and the more complex functionality will be left for users to implement. 

\subsection{Sustainability Contributions}
We identified the following two contributions to the sustainability of supply chains:
\begin{itemize}
    \item CircleChain provides a tokenization architecture and implementation that enables \textit{track} and \textit{tracing of products. This enables recycling in the context of a circular economy for more sustainable supply chains. }
    \item According to GEM \cite{gem2020} fate of $82.6\%$ of generated waste globally is uncertain. The architecture of CircleChain can be configured to add reporting capabilities that integrate information from all issued tokens to determine the fate of products.
    \item Providing an \textit{authenticator} role enables the system to ensure the quality of products as well as audit the refurbishment, reuse, and recycling of the recycler role. This helps prevent premature recycling of products.
\end{itemize}

\section{Conclusion}

We proposed a reference architecture that  facilitates the management of products in the context of a circular economy. Our system is highly scalable and secure and can be used in real-world applications. Furthermore, we have a complete open-source implementation available on our Github page \footnote[5]{https://github.com/Kasche153/CircleChain}. We also tested the scalability of the system and demonstrated how well it performed in scenarios where thousands of tokens are created and managed. 

Our system uses a two-layer solution where the lower layer manages the tokens representing the digital twin of the physical products and their state in their life cycle, whereas the upper layer manages different entities in different roles that access the tokens. Using this double-layer approach makes the whole system less prone to design and implementation flaws since the system is designed with respect to the capabilities that Algorand Standard Assets provide, not user-defined tokens with arbitrary capabilities. We have demonstrated the workflow in the context of producing and reusing chips.




\begin{table}
  \caption{Scalability test environment setup}
  \label{tab:test}
  \centering
  \begin{center}
  \begin{tabular}{p{4cm}|p{4cm}}
    \toprule
    \textbf{Experiment Parameter} & \textbf{Description} \\
    \midrule
    Test Environment & Algorand Testnet (3.5.1.stable) \\
    Synchronization & Fast catchup \\
    Node type & Normal participant node \\
    OS and hardware & Windows 11, 16 GB RAM, SSD, Intel Core i7-10700 CPU @ 2.90GHz \\
    SDK & py-algorand-sdk, asyncio \\
    \bottomrule
  \end{tabular}
  \end{center}
\end{table}

\bibliographystyle{unsrt}
\bibliography{paper}  






\end{document}